\documentclass[aps,prb,twocolumn]{revtex4-2}
\usepackage{graphicx}
\usepackage{dcolumn}
\usepackage{diagbox}
\usepackage[T1]{fontenc}
\usepackage{amssymb, amsmath,amsfonts}
\usepackage{bm}
\usepackage[usenames]{color}
\usepackage{xcolor,tabularx}
\usepackage{colortbl}
\usepackage{hyperref}

\newcommand{\ket}[1]{\left|#1\right\rangle}

\newcommand{\msem}{m}
\newcommand{\lkm}{\lambda_{KM}}
\newcommand{\lkmr}{\lambda_{KM,R}}
\newcommand{\lkml}{\lambda_{KM,L}}
\newcommand{\lkmij}{\lambda^{KM}_{ij}}
\newcommand{\lvz}{\lambda_{VZ}}
\newcommand{\lvzr}{\lambda_{VZ,R}}
\newcommand{\lvzl}{\lambda_{VZ,L}}
\newcommand{\lvzij}{\lambda^{VZ}_{ij}}
\newcommand{\lr}{\lambda_R}
\newcommand{\lrij}{\lambda^{R}_{ij}}
\newcommand{\D}{\Delta}
\newcommand{\id}{\openone}

\newcommand{\tz}{\tau_z}
\newcommand{\taus}{\tau}
\newcommand{\sx}{\sigma_x}
\newcommand{\sy}{\sigma_y}
\newcommand{\sz}{\sigma_z}
\newcommand{\sigmas}{\sigma}
\newcommand{\spx}{s_x}
\newcommand{\spy}{s_y}
\newcommand{\spz}{s_z}
\newcommand{\spins}{s}

\usepackage[utf8]{inputenc}
\usepackage[english]{babel}
\usepackage{lmodern}
\usepackage{csquotes}

\begin{document}

\title{Robust propagating in-gap modes due to spin-orbit domain walls in graphene}
   
\author{Jean-Baptiste Touchais}
\author{Pascal Simon}
\author{Andrej Mesaros}
\email{andrej.meszaros@universite-paris-saclay.fr}

\affiliation{Universit\'e Paris-Saclay, CNRS, Laboratoire de Physique des Solides, 91405, Orsay, France}

\date{\today}

\begin{abstract}
Recently, great experimental efforts towards designing topological electronic states have been invested in layered incommensurate heterostructures which form various nano- and meso-scale domains. In particular, it has become clear that a delicate interplay of different spin-orbit terms is induced in graphene on transition metal dichalcogenide substrates. {We therefore theoretically study various types of domain walls in spin-orbit coupling in graphene looking for robust one-dimensional propagating electronic states. To do so,  we use an \emph{interface Chern number} and a spectral flow analysis in the low-energy theory and contrast our results to the standard arguments based on valley-Chern numbers or Chern numbers in continuum models.} 
Surprisingly, we find that a sign-changing domain wall in valley-Zeeman spin-orbit coupling binds two robust Kramers pairs, within the bulk gap opened due to a simultaneous presence of Rashba coupling. 
We also study the robustness to symmetry breaking and lattice backscattering effects in tight-binding models.
We show an explicit mapping of our valley-Zeeman domain wall to a domain wall in gated spinless bilayer graphene. We discuss the possible spectroscopic and transport signatures of various types of spin-orbit coupling domain walls in heterostructures.
\end{abstract}

\maketitle

\section{Introduction}
\label{sec:intro}
Amid an explosion of research into van der Waals materials during this decade, graphene-based platforms remain central for discovery and design of topological states of matter. Today's promising platforms are based on inhomogeneities in real space, for example, in devices and heterostructures with designed spatial variation of order parameters and external fields. Great progress has been achieved in twisted multilayers in which new electronic states may arise in Moire patterns and domains, as observed in twisted bilayer graphene\cite{Huang18,Rickhaus18,Yoo19,Lebedeva20}. In particular, layering graphene and transition metal dichalcogenides realizes the early idea of seeking topological states in graphene by inducing spin-orbit coupling (SOC) in it \cite{Avsar14,Wang16,Zihlmann18,Wakamura19}. Experiments\cite{Yang16,Wang16,Wakamura19,Zihlmann18}, first principles calculations\cite{Cummings17,David19} and theory\cite{David19,Koshino19} agree that the outcome is complex, with at least four different induced coupling terms in accord with the lowered symmetry of the system\cite{Fabian17}: the Kane-Mele SOC, the valley-Zeeman SOC\cite{McCann12,Wang16}, the Rashba SOC, and the Dirac mass. Therefore, understanding the electronic modes due to spatial variations of multiple coupling parameters is necessary both fundamentally and practically.

A special type of electronic states designed for spin- and valleytronics involves creating one-dimensional channels, which can essentially be understood as domain walls (DW) in a certain coupling parameter.  One of the first proposals for topological channels was due to a gating DW in bilayer graphene\cite{Martin08}, which was later connected to a DW in stacking order\cite{Vaezi13}, leading to experimental observation of the one-dimensional modes\cite{Yin16}. Further DWs in Dirac mass parameter yielded valley-polarized modes\cite{Niu09,Semenoff08}. Alternative DWs using strain field as the parameter were proposed\cite{Sasaki10,Yang19}, while multilayer systems allow for even more coupling parameters\cite{SanJose13,Efimkin18,Lebedeva20,Kwan20}. A unifying viewpoint of one-dimensional topological modes as being DW modes is also exemplified in the unexpected robustness to magnetic field of quantum spin Hall helical edge states that was understood by invoking a DW between a quantum spin Hall and a quantum Hall domain\cite{Shevtsov12}. Understanding DWs in presence of multiple parameters is crucial, as becomes clear in the recent finding that quantum spin Hall edge modes may be transformed or supplemented by spin- or valley-polarized modes as one changes the dominant SOC parameter\cite{Frank18,Fabian17}. Nevertheless, the treatment of non-randomly varying SOC parameters in this context remains scarce\cite{Hosseini15,Brataas07}.

{Here we mainly focus on domain walls in valley-Zeeman and Rashba spin-orbit couplings in graphene, and contrast them to known domain walls in Dirac mass and Kane-Mele SOC. Our main finding is that a domain-wall in valley-Zeeman SOC in presence of arbitrary constant Rashba SOC hosts two valley-polarized Kramers pairs propagating along the domain wall. Importantly, we find that these modes are not protected by a bulk topological index of the 10-fold way \cite{Schnyder2008,Kitaev2009,Chiu16}, nor an index derived from the remaining lattice symmetry ($C_{3v}$), but instead by an ``interface Chern number''. In essence, the bulk ``valley Chern number'' index $C_{bulk}^\tau$, i.e., the Chern number calculated in the graphene continuum model for a fixed valley index $\tau=\pm1$, has opposite values on the two domains with opposite signs of valley-Zeeman SOC, but this continuum-derived bulk index is ill-defined and does not provide topological protection of modes even without any intervalley scattering, any symmetry breaking, nor
bulkgap closing \cite{MartinMarginal}. Alternatively, it was proposed that the \emph{difference} of two non-zero values $\delta C_{bulk}^\tau=C_{bulk,R}^\tau -C_{bulk,L}^\tau$, occurring at a \emph{domain wall}
interface between the right (R) and left (L) bulk, is topologically well-defined \cite{MartinMarginal}. However, we show explicitly that $\delta C_{bulk}^\tau$ incorrectly predicts modes in general, and hence does not offer a ``bulk-interface" correspondence. In contrast, we find that an ``interface Chern number'' $C_{interface}^\tau=2\tau$ strictly predicts the valley-Zeeman DW modes by applying a more general spectral flow theorem for Berry-Chern monopoles due to varying parameters \cite{DelplaceNotes}. The theorem is an exact statement about the existence of quantum chiral domain-wall-bound modes due to a topological index associated to a degeneracy point of the bands of an auxiliary homogeneous Hamiltonian. This degeneracy point corresponds to the real-space point where valley-Zeeman SOC changes sign, and is hence fully determined by the domain-wall and not by the topology of the domains.}

{Our work is one demonstration of the usefulness of the spectral flow theorem for topological modes in inhomogeneous quantum problems, for which there is a limited number of methods.} One may treat spatial coordinates classically at long distances from a topological defect, and consider the resulting band topology due to discrete symmetries\cite{Teo10,Chiu16}. One may also use real-space expressions that give a local indication of non-trivial topology\cite{Prodan10}. Nevertheless, a precise ``bulk-defect'' correspondence in this case remains quite abstract and invokes generalized bulk topological numbers\cite{Prodan09,Sheng06} to cause some DW modes\cite{MartinMarginal,Ezawa13,Ezawa13si}. In contrast, the spectral flow theorem\cite{Faure19,Delplace20,DelplaceNotes} builds on the notion of using topological numbers associated to local information in parameter space\cite{Volovik11,Martin08}. {In our case, the valley-Zeeman domain-wall is an interface across which parameters of a Dirac equation vary in real-space, and the theorem is useful both conceptually and practically as for our Dirac operator we do not have a standard index theorem, while the direct solution for the in-gap modes spectrum is much more tedious and opaque.}


{The uncovered valley-Zeeman DW modes are in contrast to previously identified graphene edge modes due to valley-Zeeman SOC, since the latter are fragile to the Rashba SOC strength,\cite{Frank18} and to some lattice terminations even in absence of intervalley scattering\cite{MartinMarginal}, due to the modes being connected only to the bulk valley-Chern number $|C_{bulk}^\tau|=1$\cite{Frank18,Yang16,Alsharari16}. Instead, we find an exact mapping of our valley-Zeeman SOC DW modes onto modes of a DW between two electrically gated regions in spinless bilayer graphene,\cite{Martin08,Klinovaja12} which were experimentally detected\cite{Yin16,Vaezi13}, but whose protection by $\delta C_{bulk}^\tau$\cite{MartinMarginal} is in this work shown to rather be due to $C_{interface}^\tau$. Since experiments indicate that proximitized graphene acquires spatially-dependent SOC of all three types, where valley-Zeeman is considerable\cite{Yang16,Wakamura19,Zihlmann18}, we expect that our valley-Zeeman domain-wall modes may contribute to spectroscopic and transport properties.}

This paper is organized as follows: {In Section~\ref{sec:secChern} we start by motivating the use of an interface Chern number, and discuss its connection to a difference of bulk Chern numbers.} In Section~\ref{sec:secvzDW} we first define domain walls in continuum theory of graphene, we apply the spectral flow theorem to DWs in valley-Zeeman SOC with a constant Rashba SOC, and show the mapping to spinless bilayer graphene. Then we introduce domain walls in the tight binding lattice model of graphene and discuss the robustness of valley-Zeeman DW modes, {as well as of modes on other domain walls. We finish with a discussion of potential impact on experiments, other domain walls in Dirac mass and Kane-Mele SOC, and an outlook. Technical details supplementing our analysis appear in four appendices.}

\section{The interface Chern number}
\label{sec:secChern}
\begin{figure*}
	\centering
	\includegraphics[width=0.8\textwidth]{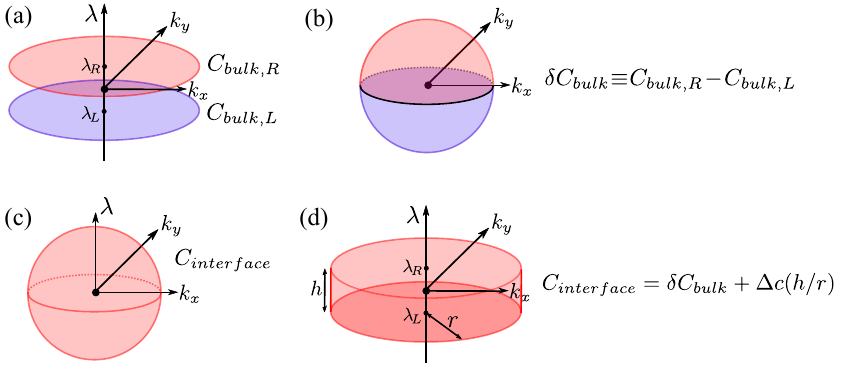}
	\caption{\label{fig:Chern} (a) Two planar integrals in the continuum giving the bulk Chern numbers associated with the left (L) and right (R) domain. These Chern numbers are unstable and the bulk-edge correspondence for either R or L is non-universal. (b) When the difference of two Chern numbers      $\delta C_{bulk}$ is considered, as seems natural for a domain wall, one may compactify the planes in (a) and obtain a stable bulk index, which however is only handwavingly connected to domain wall modes. (c) Sphere integration defining the interface Chern number $C_{interface}$. (d) We deform the sphere into a cylinder, and as obvious in comparison to panel (a), one finds that $C_{interface}=\delta C_{bulk}+\Delta c$, where $\Delta c$ may be a function of the aspect ratio $h/r$.}
\end{figure*}
{In this work we apply the general theory of spectral flow due to Berry-Chern monopoles\cite{DelplaceNotes,Faure19,Delplace20}, i.e., a spectral flow theorem (SFT), to the particular case of one spatially varying parameter in a two-dimensional system, i.e., a domain wall profile of a SOC in graphene. The application of SFT is presented in full detail in Appendix~\ref{app:SFT}, while here we sketch its form and relationship to bulk topology.}

{We start from the bulk topology. Importantly, for smooth domain walls (and we discuss sharp ones in Sec.~\ref{sec:tbDW}), a priori the intervalley scattering is negligible and the pertinent analysis is of the continuum model of graphene with valley index $\tau$ conserved. The essential problem of valley-based bulk topology was exposed in Ref.~\onlinecite{MartinMarginal}, which deals with the model of bilayer graphene with a DW profile of gate voltage --- the model which we show maps (in its spinless version) exactly to our model of valley-Zeeman DW in graphene with constant Rashba SOC (see Section~\ref{sec:secBilayer}). In a nutshell they show that the bulk ``valley Chern number'' index $C_{bulk}^\tau$, i.e., the Chern number calculated in the continuum model for a fixed valley index $\tau=\pm1$, is actually not protecting an edge mode. The key reason is that any Chern number calculated for the infinite $(k_x,k_y)$ plane (i.e., in the single-valley continuum model) is not topologically well-defined\cite{DelplaceNotes}. Hence, the edge modes disappear depending on microscopic details of the edge, even without any intervalley scattering, any symmetry breaking, nor bulkgap closing.}

{The $C_{bulk}^\tau$ itself obviously cannot predict DW modes either, but Ref.~\onlinecite{MartinMarginal} argues that in contrast the \emph{difference} of two non-zero values $\delta C_{bulk}^\tau=C_{bulk,R}^\tau -C_{bulk,L}^\tau$, occurring at a \emph{domain wall} interface between the right (R) and left (L) bulk, is topologically well-defined. Hence their gate-voltage DW modes in bilayer graphene are theoretically indeed stable. This argument is illustrated in Fig.~\ref{fig:Chern}a,b, with the key point that the bandstructure of R and L domains becomes equal far away from $(k_x,k_y)=(0,0)$, so in the difference $\delta C_{bulk}^\tau$ one may compactify the two planes into a sphere. Based on the exact mapping between the bilayer gate-voltage DW problem and our valley-Zeeman DW problem, it seemingly follows that the difference $\delta C_{bulk}^\tau$ provides protection to our modes too.}

{To the contrary, our point is that the $\delta C_{bulk}^\tau$ is also fragile in general, while the interface Chern number $C_{interface}^\tau$ and the SFT is generally well-defined and more powerful for interface (domain-wall) problems. Namely, in this context, (1) A $C_{interface}$ can exist without reference
to any well-defined bulk quantity or compact parameter space, which are absent in valley-resolved situations, and (2) The SFT is an exact statement about the quantum domain-wall modes, as opposed to a bulk-boundary correspondence which is vague or it is  justified \emph{a posteriori} by
tedious explicit calculations. We now consider the relationship between $C_{interface}^\tau$ and $\delta C_{bulk}^\tau$ more explicitly.}

{Let us now introduce the $C_{interface}^\tau$, assuming that a Hamiltonian parameter $\lambda$ (such as SOC) varies smoothly along the $x$-axis (as in a domain-wall profile $\lambda(x)$): the SFT prescribes (see Sec.~\ref{app:SFT} and Ref.\onlinecite{DelplaceNotes} for details) that we look at an auxiliary Hamiltonian matrix $\tilde{H}$ in which the operator $-i\partial_x$ is replaced by a real parameter $k_x$, and that we seek band degeneracies of $\tilde{H}$ as function of $\lambda(x), k_x, k_y$. If the interface profile causes such a degeneracy at the parameter values $(x,k_x,k_y)=(x_c,k_x^c,k_y^c)\equiv p_c$, one needs to calculate the interface Chern number $C_{interface}^\tau$ of filled bands of $\tilde{H}$ on a sphere surrounding $p_c$. Hence the naming of ``interface'' Chern number, which uses only information from the vicinity of the point $p_c$ and does not involve any integrations over, e.g., a Brillouin zone.} 

{With this definition of $C_{interface}^\tau$, in Fig.~\ref{fig:Chern}c,d we devise a procedure where the integration in $C_{interface}^\tau$ is smoothly deformed from a sphere to a cylinder to give 
\begin{equation}
  \label{eq:5}
C_{interface}^\tau=\delta C_{bulk}^\tau+\Delta c^\tau,  
\end{equation}
where one takes the limit for the radius of the cylinder $r\rightarrow\infty$; the difference $\Delta c^\tau\equiv\lim_{h/r\rightarrow0}\Delta c^\tau(h/r)$ is given by the Chern integral on the side of cylinder keeping its height $h$ finite (could be arbitrarily small as long as the cylinder encloses the
degeneracy point of $\tilde{H}$ at the origin). For our VZ DW continuum model we explicitly find $\Delta c^\tau=0$. Hence in our model, $C_{interface}^\tau$ and $\delta C_{bulk}^\tau$ coincide. However, note that
only the $C_{interface}^\tau$ invariant guarantees the domain-wall modes through the SFT, while the exact correspondence between $\delta C_{bulk}^\tau$ and domain-wall modes is not guaranteed and may be found \emph{a posteriori} (e.g., Ref.~\onlinecite{MartinMarginal} had to solve explicitly for the modes). 

In order to make this point concrete, we consider a simple continuum model (with a single valley) in which the $C_{interface}=1$ protects a chiral domain-wall mode, while in contrast the corresponding $C_{bulk}=0$, and hence $\delta C_{bulk}=0$, so the bulk-derived Chern numbers predict the absence of modes on edges and domain walls. In this example the interface Chern number comes entirely from the side of the cylinder in Fig.~\ref{fig:Chern}d, while the cylinder bases contribute zero, i.e., $\Delta c=1$. The model is based on a single spin component of the Bernevig-Hughes-Zhang model with a gapped Dirac cone at the $\Gamma$ point:
\begin{equation}
  \label{eq:1}
  H_{bulk}=k_x\sigma_x+k_y\sigma_y+\lambda(M-k_x^2-k_y^2)\sigma_z,
\end{equation}
where we will be imagining a DW in the parameter $\lambda(x)$ going from $\lambda_L\equiv\lambda(x=-\infty)=+1$ to $\lambda_R\equiv\lambda(x=+\infty)=-1$, while keeping $M<0$ constant. Thinking of either of the two bulk domains, $\lambda(x)\equiv\lambda_L$ or $\lambda(x)\equiv\lambda_R$, we know that for any non-zero $\lambda$ the $M<0$ bulk Hamiltonian is trivial and $C_{bulk}(\lambda=\pm1)=0$, and obviously $\delta C_{bulk}=0$. In contrast, for any $M<0$ we find $C_{interface}=1$ due to the degeneracy point $(k_x,k_y,\lambda(x))=(0,0,0)$ and hence there is one chiral mode on a DW where $\lambda(x)$ changes sign. We confirm this in a numerical solution. Looking at the cylinder in the limit $h/r\rightarrow 0$ we find explicitly that in this model the $C_{interface}=\Delta c$ comes entirely from the side of the cylinder.}

{We will discuss the robustness of modes protected by $C_{interface}$ in Sec.~\ref{sec:lattVZDW}, after we introduce the lattice model and intervalley scattering.}

\section{Domain walls in valley-Zeeman SOC}
\label{sec:secvzDW}
\begin{figure}
	\centering
	\includegraphics[width=0.45\textwidth]{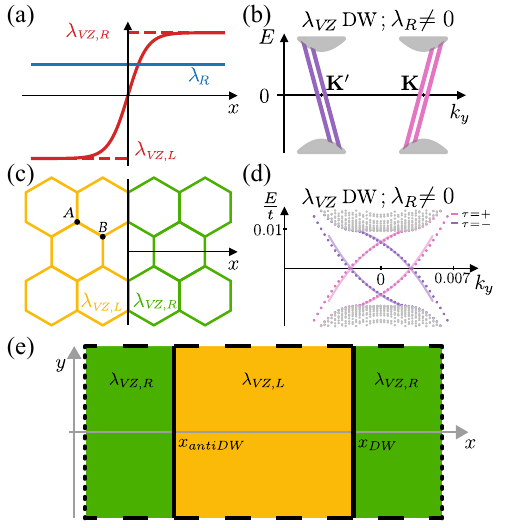}
	\caption{\label{fig:1} Modes of a domain wall in valley-Zeeman (VZ) SOC, and their robustness on the lattice. (a) Generic profile of sign-changing DW in VZ SOC $\lvz$, with arbitrary non-zero value of Rashba SOC $\lr$. (b) Sketch of modes in momentum space predicted by the spectral flow theorem for (a). Both valleys $K,K'$ are projected to $k_y=0$, but are offset here for visibility. (c) An atomically sharp profile of an armchair DW on the lattice, corresponding to (a). (d) Tight-binding lattice model spectrum for the system sketched in (e), with parameters $N_x=801$, $N_y=2000$, $\lvz(\pm\infty)=\pm 0.01t$ and $\lr=0.1t$. In the spectrum we only show the in-gap modes localized around the DW located at $x_{DW}$ (see (e)), with purple/pink being states in valley $K/K'$ (both valleys project to $k_y=0$ on the lattice). The modes match the spectrum of DW in gating of bilayer graphene (full lines; see text). (e) Sketch of the lattice with periodic boundary conditions (identifying two dashed edges with each other, and likewise the dotted) and consequently a DW at position $x_{DW}$ and an ``anti-DW'' of opposite orientation at $x_{antiDW}$.}
\end{figure}
\subsection{General model of smooth domain walls in the continuum}
\label{sec:smoothDW}
We start by considering the low-energy continuum Dirac theory for graphene with a smooth domain wall in any of our coupling parameters:
\begin{align}
  \label{eq:Hcont}
  H=&\tz\sx(-i\partial_x)+k_y\sy+\msem(x)\sz+\lkm(x)\tz\sz\spz \notag\\ 
  &+\lvz(x)\tz\spz+\lr(x)(\tz\sx\spy-\sy\spx),
\end{align}
where we assume translational invariance along the domain wall in $y$ direction (any direction is equivalent in the low-energy theory), the $\msem$, $\lkm$, $\lvz$ and $\lr$ are the Dirac mass, Kane-Mele spin orbit (KM), valley-Zeeman spin-orbit (VZ) and Rashba spin-orbit (R) couplings, respectively, and the Pauli matrices $\taus_i$, $\sigmas_i$, $\spins_i$, $i=x,y,z$, are acting in the valley, sublattice and spin space, respectively, while we set $\hbar v_F\equiv1$. Throughout the paper for simplicity we consider coupling parameters in pairs: one having a DW profile (detailed below), the other one being constant and non-zero, while the rest of coupling parameters are set to zero. We further require that the two domains far away from the DW are gapped, so that any DW modes are prominent inside a bulkgap. Obviously, if multiple well-separated DWs appear in the system, they will behave independently and each will carry its modes. In this section we focus on the new DW in $\lvz(x)$ with $\lr$ constant. In Appendix~\ref{app:restDWs} we show that a DW in $\lr(x)$ with $\lvz$ constant does not carry topological modes, while for DWs in $\lkm(x)$ and $\msem(x)$ we recover their already known DW modes.

The profile of a domain wall in any one of the couplings $\D\in\{\msem,\lkm,\lvz,\lr\}$ is defined as:
\begin{equation}
  \label{eq:DWcont}
  \D(x)\equiv\frac{\D_L+\D_R}{2}+\frac{\D_R-\D_L}{2}\eta(x),
\end{equation}
characterized by the limiting values $\D_L\equiv\D(x\rightarrow-\infty)$, $\D_R\equiv\D(x\rightarrow\infty)$, which define the domains to the left and to the right of the DW, and by an arbitrary smooth function $\eta(x)$, which satisfies $\eta(x\rightarrow\pm\infty)=\pm1$ (see Fig.~\ref{fig:1}a). For a single DW the profile $\eta(x)$ may be taken as monotonous, and its particular form has no bearing on the results in this section: one may imagine a typical profile such as $\eta(x)\in\{\frac{2}{\pi}\arctan(x/l), \tanh(x/l)\ldots\}$ with a finite width length-scale $l$. We only consider DW modes in cases where both domains $L,R$ have a full gap far away from the DW.

\subsection{Valley-Zeeman domain walls in the continuum and the interface Chern number}
\label{sec:smoothVZDW}
We use the spectral flow theorem to show that a DW in valley-Zeeman spin-orbit coupling in presence of a constant non-zero Rashba spin-orbit coupling hosts propagating modes which were not identified before. First, a domain with constant $\lvz,\lr\neq0$ has a gap $\frac{2|\lr\lvz|}{\sqrt{\lr^2+\lvz^2}}$ at half filling even though neither coupling on its own opens a gap in graphene. Hence we consider a domain wall in $\lvz(x)$. The spectral flow theorem, whose application is presented in full detail in Appendix~\ref{app:SFT}, prescribes to look at an auxiliary Hamiltonian matrix $\tilde{H}$ in which $-i\partial_x$ is replaced by a parameter $k_x$, giving the spectrum:
\begin{widetext}
\begin{equation}
  \label{eq:Evzr}
  E_{\taus,\alpha\beta}=\alpha\sqrt{k^2+\lvz(x)^2+2\lr^2+2\beta\sqrt{\lr^4+(\lr^2+\lvz(x)^2)k^2}},
\end{equation}
\end{widetext}
with $\alpha,\beta=\pm$, $k^2=k_x^2+k_y^2$, having a valley degeneracy ($\taus=\pm1$) of each band. As noted, this spectrum has a gap if $\lvz,\lr\neq0$. We now look at gap-closing degeneracy points at half filling, i.e., at band-touching points between the two inner bands, which are the ones having $\alpha=\pm1$ and $\beta=-1$. At a constant $\lr\neq0$ the only such gap-closing may happen at $(x,k_x,k_y)=(x_c,0,0)$ such that $\lvz(x_c)=0$. With this degeneracy point identified, the spectral flow theorem prescribes (see Appendix~\ref{app:SFT} for details) that we construct the 4x4 projector (in Hilbert space of $\sigmas,\spins$) to the 2 filled bands of $\tilde{H}$, $P_\taus=P^{\alpha=-,\beta=+}_\taus+P^{\alpha=-,\beta=-}_\taus$, and we do it using the formula for the projector to the $n$-th band: $P^{(n)}=\prod\limits_{m\neq n}\frac{\tilde{H}-E_m}{E_n-E_m}$. Finally, in the parameter space $(\lvz(x),k_x,k_y)$ we take a small sphere $S^2$ enclosing the degeneracy point $(0,0,0)$, and evaluate the Chern number of $P_\taus(\lvz(x),k_x,k_y)$ on this sphere, obtaining the result
\begin{equation}
  \label{eq:3}
C_\taus=2\taus\,\text{sgn}(\lvzr-\lvzl).
\end{equation}

The outcome of the spectral flow theorem is that a DW across which $\lvz$ changes sign, in presence of any non-zero $\lr$ as shown in Fig.~\ref{fig:1}a, has $|N^\pm|=2$ chiral propagating modes with the direction of velocity along $y$ being $\text{sgn}(N^\pm)=\pm1$ in each valley $\taus=\pm$ (Fig.~\ref{fig:1}b).

\subsection{Mapping to spinless bilayer graphene}
\label{sec:secBilayer}
The valley-Zeeman DW modes can be exactly mapped onto the bound states of a DW profile in gate voltage in spinless bilayer graphene introduced in Ref~\cite{Martin08}. The low energy Hamiltonian of AB-stacked gated bilayer graphene is:
\begin{equation}
H_{BLG}=k_x\sx + k_y\tz\sy+\frac{t_\perp}{2}(\sigma_x\eta_x+\sigma_y\eta_y)-V\eta_z,
\end{equation}
where the new Pauli matrices $\eta_i$ act in the layer space, while $t_\perp$ is the hopping amplitude to go from the $A$ atom of one layer to a $B$ atom of the other layer. The mapping to our graphene model with valley-Zeeman and Rashba can be done stepwise by applying two unitary transformations. First, to recover our form of the kinetic part we apply $U^{\dagger}=\frac{1+\tz}{2}+\frac{1-\tz}{2}\sz$ which multiplies the $\tau=-1$ sector by $\sz$, obtaining
\begin{equation}
H^{(1)}_{BLG}=k_x\tz\sx+k_y\sy+\frac{t_\perp}{2}\tz(\sx\eta_x+\sy\eta_y)-V\eta_z
\end{equation}
Second, we map the layer exchange part onto Rashba SOC, and the gate voltage onto valley-Zeeman SOC by performing a rotation around $\eta_z$ to exchange $x$ and $y$, and then multiplying by $\eta_y$ in the $\tau=-1$ sector, which altogether requires $U^{\dagger}=\frac{1-i\eta_z}{\sqrt{2}}(\frac{1+\tz}{2}+\frac{1-\tz}{2}\eta_y)$, so that in the end we have
\begin{equation}
H^{(2)}_{BLG} =k_x\tz\sx+k_y\sy+\frac{t_\perp}{2}(\tz\sx\eta_y-\sy\eta_x)-V\tz\eta_z.
\end{equation}
By reinterpreting the spinless bilayer's layer degree of freedom as a spin degree of freedom we finally get the Hamiltonian of graphene with a valley-Zeeman SOC $\lvz=-V$ and a Rashba SOC $\lr=\frac{t_\perp}{2}$.

In the limit where $V\ll t_{\perp}\Rightarrow\lvz\ll\lr$ which was studied in Ref.~\cite{Martin08} they show that for $V(x)=\kappa V_0\mathrm{sgn}(x)$ with $t_\perp>0$, $V_0>0$, and $\kappa=\pm 1$, there are 4 in-gap branches crossing zero energy:
\begin{equation}
E_{\tau\pm} = \pm\left(\frac{v_Fk_y\tau\kappa}{2\sqrt{t_\perp}}\mp\sqrt{\frac{v_F^2k_y^2}{4t_\perp}+\frac{V_0}{\sqrt{2}}}\right)^2\mp\sqrt{2}V_0.
\end{equation}
The crossings appear at $k_y=\mp\frac{\kappa}{v_F}\sqrt{\frac{t_\perp V_0}{2\sqrt{2}}}$, while the sign of the velocity around the crossings is given by $-\kappa\tau$. Given the identification $\lvz(x)=-V(x)$, it means that in our model for a DW with negative values of valley-Zeeman SOC on the left and positive values on the right, the bound states with $\tau=1$ are right movers, which matches our results, see Fig.~\ref{fig:1}a,b.

\subsection{General model of domain walls on the lattice}
\label{sec:tbDW}
In the preceding continuum theory the electronic modes hosted by DWs are labeled by a valley index $\taus$,
so even the spectral flow theorem cannot prevent the mixing and gapping-out of modes in presence of inter-valley scattering on the graphene lattice. Therefore we study via exact diagonalization the tight-binding lattice models of DWs with two main goals: (i) To confirm the continuum theory predictions when DW profile varies slowly over many lattice sites; (ii) To assess the robustness to intervalley scattering and lattice anisotropy using an atomically sharp DW profile.

The tight-binding Hamiltonian collecting all the position-dependent coupling terms we consider is:
\begin{align}
  H =& -t\displaystyle\sum_{<i,j>}c_i^\dagger c_j+\sum_{i,\alpha} (-1)^{l_i}\msem_{i}c_{i\alpha}^\dagger c_{i\alpha}\\\notag
  &+ i\sum_{<<i,j>>,\alpha,\beta}\lkmij\nu_{ij}c_{i\alpha}^\dagger \spz^{\alpha\beta} c_{j\beta} +H.c. \\\notag
        &+ i\sum_{<<i,j>>,\alpha,\beta}\lvzij (-1)^{l_i}\nu_{ij}c_{i\alpha}^\dagger \spz^{\alpha\beta} c_{j\beta}+H.c.\\\notag
        &+i\sum\limits_{\substack{<ij>,\alpha,\beta}}\lrij\,\hat{z}\cdot(\vec{d}_{ij}\times\vec{\spins}^{\alpha\beta})\, c^\dagger_{i\alpha}c_{j\beta} +H.c.,
	\label{eq:lattice_Hamiltonian}
\end{align}
where $c^\dagger_{i\alpha}$ creates an electron on site $i$ with $S_z$ spin $\alpha=\pm$, the $(-1)^{l_i}=+(-)$ for a site on sublattice $A(B)$, the $\nu_{ij}=\textrm{sgn}[\hat{z}\cdot\left(\vec{d}^{(1)}_{ij}\times\vec{d}^{(2)}_{ij}\right)]$ with $\vec{d}^{(1)}_{ij},\vec{d}^{(2)}_{ij}$ the two NN-bond vectors forming the path $j\rightarrow1\rightarrow2\rightarrow i$ between NNN neighbors $j,i$, where we normalize $|\vec{d}_{ij}|=1$. When the couplings are homogeneous, independent of sites $i,j$, one finds the quantitative connection to the continuum couplings, i.e., $\msem_i=\msem$, $\lkm=-3\sqrt{3}\lkmij$, $\lvz=-3\sqrt{3}\lvzij$, $\lr=-3\lrij/2$.

{Before introducing the models for domain walls, let us recall the crystalline symmetries of the various spin-orbit terms we introduced in the above Hamiltonian. The Kane-Mele SOC preserves the full graphene point-group symmetry $D_{6h}$. The valley-Zeeman SOC breaks symmetries exchanging sublattices, such as inversion and $C_6$, but not the $z\rightarrow-z$ mirror symmetry, hence preserving the point group $D_{3h}$. The Rashba SOC breaks the $z\rightarrow-z$ mirror symmetry and inversion so preserves $C_{6v}$. Finally, the Dirac mass preserves the $D_{3h}$ symmetry such as the valley-Zeeman SOC.}

Now we consider a graphene lattice with two domains. Requiring periodic boundary conditions on the lattice forces the existence of two domain walls since the domains meet each other twice, see Fig.~\ref{fig:1}e. More precisely, consider first the profile of the coupling constant which creates one DW on the lattice:
\begin{equation}
  \label{eq:4}
\eta(\vec{R}_i,\vec{R}_0)=\textrm{tanh}\left[\frac{(\vec{R}_i-\vec{R}_0)\cdot\vec{e}_{DW}}{l}\right],  
\end{equation}
where $\vec{R}_i$ is the position of site $i$, $\vec{R}_0$ is a position in the center of a honeycomb plaquette through which the DW passes, while the $\vec{e}_{DW}=\hat{x}(\hat{y})$ gives an armchair(zigzag) DW on the lattice, see Fig.~\ref{fig:1}c. The Hamiltonian has translational symmetry along the straight DW, and consequently the bulk Dirac points are projected onto two distant momenta $k_x$ in case of zigzag DW, and onto the same $k_y=0$ in case of armchair DW. Effects of intervalley scattering are consequently masked in the zigzag DW case, and we find the predicted continuum modes. Therefore in the rest of the paper we present the armchair DWs for which the intervalley lattice scattering effects are fully exhibited. The length $l$ is used to vary the smoothness of the DW profile and therefore tune the amount of intervalley scattering. By ``sharp DW'' we mean the limit $l\rightarrow 0$, where $\eta(\vec{R}_i,\vec{R}_0)$ becomes a step-function (Fig.~\ref{fig:1}c). For a sharp DW we verify that the precise value of couplings on NNN bonds which cross the domain boundary do not matter for the main features of in-gap modes.

To satisfy the periodic boundary conditions as in Fig.~\ref{fig:1}e we position the DW to cross $\vec{R}_0\equiv\vec{R}_{DW}$ and overlay a second domain wall profile (the so-called ``anti-DW'') to cross $\vec{R}_0\equiv\vec{R}_{antiDW}$:
\begin{align}
  \label{eq:lattDW}
&\eta(\vec{R}_i)\equiv\eta(\vec{R}_i,\vec{R}_{DW})\cdot\eta(\vec{R}_i,\vec{R}_{antiDW}),\\
&\D_i\equiv\frac{\D_L+\D_R}{2}+\frac{\D_R-\D_L}{2}\eta(\vec{R}_i).
\end{align}
The sketch in Fig.~\ref{fig:1}e represents an armchair DW and anti-DW, hence their positions are given by $x_{DW}=\vec{R}_{DW}\cdot \hat{x}$ and $x_{antiDW}=\vec{R}_{antiDW}\cdot \hat{x}$. The coupling $\D$ can be any one of $\D\in\{m,\lkm,\lvz,\lr\}$, and by construction on the lattice $\D_i$ has the values $\D_L(\D_R)$ on the left(right) side of the DW, but has the opposite values $\D_R(\D_L)$ on the left(right) side of the anti-DW (hence the name ``anti''), see Fig.~\ref{fig:1}e for the example of valley-Zeeman SOC, $\D\equiv\lvz$.
Since the bulkgap is given by the smaller of the bulkgaps on two domains, for simplicity our profile has $-\D_L=\D_R\equiv \D$.

Electronic states localized in real space around DW and around anti-DW are degenerate by lattice symmetry. To identify these states separately, we energetically split them by adding a small chemical potential $\sum_\alpha c_{i\alpha}^\dagger c_{i\alpha}$ in real space at the lattice sites $i$ along the DW. To assign a valley index to an electron state, we Fourier transform its $x$-dependence, and note that the two valleys contribute Fourier components at opposite momenta $\pm K$.

\subsection{Robustness of Valley-Zeeman domain-wall modes}
\label{sec:lattVZDW}
The main result of the tight-binding lattice model with an armchair DW profile in valley-Zeeman SOC $\lvzij$ and a constant non-zero Rashba SOC $\lrij$ as defined in the previous subsection, is that we find two co-propagating modes in each valley per DW in exact accord with the new modes identified in the continuum in Section~\ref{sec:smoothVZDW}. 

{We are now in a position to discuss the robustness of these valley-Zeeman DW modes provided by the interface Chern number $C_{interface}^\tau=2\tau$ and the spectral flow theorem.  We study three ways to destroy the DW modes: 
(1) Intervalley scattering, which destroys the $\tau$ number, and hybridizes the modes in the same way as it would any modes protected by some bulk topology $C_{bulk}^\tau$ in general.
(2) Closing the direct bulkgap, which removes a prerequisite in the SFT proof of domain-wall modes, is as detrimental as for bulk-topology protected modes. 
(3) Destroying translation symmetry along the domain-wall, which figures explicitly in SFT and is hence a peculiarity of protection by $C_{interface}$. 
}

{First, we consider closing the direct bulkgap. We add a constant $\lkmij$ coupling, which is a natural choice since all three types of spin-orbit coupling appear in graphene on transition metal dichalcogenide substrates\cite{Yang16,Wakamura19,Zihlmann18,David19,Koshino19}. The in-gap DW modes indeed stay gapless with increasing $\lkm$ all the way until the bulkgap opened by $\lvzij$ and $\lrij$ closes on both domains. This is fully explained in the SFT picture, since once the $\lkmij$ closes and reopens the bulkgap in competition with the valley-Zeeman\&Rashba, the degeneracy point in parameter space disappears.\footnote{{This simple effect of competing bulkgaps (without any DWs) between Kane-Mele SOC and valley-Zeeman and Rashba SOC was already identified in Refs.~\cite{Frank18,Fabian17}.}} Note, the SFT does not require a bulk insulator to stabilize the DW modes, since it relies on the bandstructure only in the vicinity of the degeneracy point in parameter space. In other words, the bulkgap could close at some far away point in the Brillouin zone of the domains without affecting the DW modes. In contrast, bulk topological numbers would not even be defined in such a scenario.}

{Second, we consider intervalley scattering. Within the energy resolution of our tight-binding model, the gapless DW modes persist even for a sharp DW profile, see Fig.~\ref{fig:1}d (compare to continuum result in Fig.~\ref{fig:1}b). This is however a consequence of the fact that SOC is a weak intervalley scatterer, so although the atomically sharp armchair DW has large intervalley scattering Fourier amplitudes, these are suppressed by a geometric scattering prefactor of the SOC. We confirmed that the valley-Zeeman DW modes can be gapped by adding strong enough atomic-scale impurities, as expected, so that this robustness is only parametrical. Let us compare this behavior to the well-known valley-polarized modes for a sign-changing DW of Dirac mass $\msem$, which are known to be sensitive to intervalley scattering\cite{Semenoff08}. We are able to quantify the effect. Treating a sharp DW as a perturbation to ideal graphene (see details in Appendix~\ref{app:mass}), we find that intervalley scattering at second order of perturbation opens a gap of size $2\frac{2\msem^2}{3t}$ in the DW modes, which matches very well the tight-binding results, Fig.~\ref{fig:app2}a,b. Our perturbative approach is applicable only if $\msem<t$, which translates to the gapped DW modes still remaining inside the bulkgap. Interestingly, even a small smoothing of the DW profile drastically reduces this intervalley scattering effect, e.g., a DW profile with width of a few lattice constants $l\sim2a$ reduces the gap in the modes by an order of magnitude, see Fig.~\ref{fig:app2}c.}

{Third, destroying translation symmetry along the DW, i.e., removing the $k_y$ quantum number of the armchair DW, which figures explicitly in the SFT and is hence a peculiarity of protection by $C_{interface}^\tau$. As one way to test this, in our tight-binding model of the valley-Zeeman DW, we added a random component to the VZ SOC in wide strips covering each DW, and observed that even with appreciable random component (with standard deviation of order of the bulkgap) the density of states inside the bulkgap remains the same as for the perfect DW modes, while we check that the localization of DW modes does not change appreciably. Hence there is no pathological sensitivity to deforming the translationally invariant DW.}

We close this section with a detailed comparison of our valley-Zeeman DW modes and the previously identified\cite{Frank18} zigzag-ribbon edge modes of graphene with $\lvz$ in the so-called ``quantum valley spin Hall state" (QVSHS). The homogeneous bulk model used to study QVSHS is identical to our homogeneous bulk model, when they both focus on graphene with valley-Zeeman and Rashba SOC.

We consider \emph{zigzag} edges and DWs to eliminate intervalley scattering effects and focus on topological protection with $\tau$ fixed. In terms of topological protection, we have two co-propagating modes \emph{per domain wall} in each valley due to $|C_{interface}^\tau|=2$, while the QVSHS has one ``valley-centered'' mode \emph{per edge} in each valley associated to the (topologically ill-defined) valley-Chern number $|C_v|=1$ \cite{Frank18,Yang16,Alsharari16}. To prove the different degrees of topological protection, in our tight-binding model we make the hoppings located on the domain walls tunable, so that we can smoothly interpolate between a system with two zigzag DWs (separating two domains in a periodic lattice) and a system with four zigzag edges (when the two domains become disconnected) \footnote{{The additional ``pseudohelical edge modes'' of QVSHS connect valleys, so they cannot be matched to our DW modes, and indeed these modes are removed from the low-energy part of the spectrum as four edges evolve into two DWs.}}. We find that the four valley-centered QVSHS edge modes evolve into the four valley-Zeeman DW modes. Strikingly, the QVSHS edge modes are gapped out by increasing the value of $\lr/\lvz$ consistent with the claim in Ref.~\onlinecite{Frank18}, although this does not change the band topology. In contrast, the DW modes remain gapless, as predicted by the SFT and $C_{interface}^\tau$ for any value $\lr\neq0$.\footnote{{Here we note that Ref.~\onlinecite{MartinMarginal} considers different zigzag/bearded edge terminations in each layer of their bilayer graphene model, and find that they might have 0,1, or 2 valley-centered modes per edge per valley in accordance with fragility of $C_{bulk}^\tau$. However, the mapping of their model to ours that we discovered takes their layer index into our $S_z$ index, hence the only natural edge terminations in our model could be both zigzag or both bearded zigzag, and these options each have 1 mode per edge per valley. Hence, we demonstrated the fragility of edgemodes using the strength of Rashba SOC, instead of using a switch between zigzag/bearded.}}

\section{Discussion and Conclusions}
Using the spectral flow theorem we derived topologically protected electronic modes of various domain walls in graphene with Dirac mass, Kane-Mele SOC, valley-Zeeman SOC and Rashba SOC, with precise symmetry and chirality labeling. However this method does not address the robustness to breaking symmetries and to lattice effects, for which we employed tight-binding modeling.

{The main finding is} the robust pair of Kramers pairs on a valley-Zeeman DW in presence of a constant Rashba SOC, which might be relevant to the efforts of inducing topological phases in graphene by proximity to transition metal dichalcogenides; namely, experiments find that the induced valley-Zeeman SOC is strong, and there is a weaker Rashba SOC, at least on a large-scale average\cite{Wakamura19,Zihlmann18}. Due to incommensurability, on the scale of Moire pattern there can be domains where couplings vary significantly and even change sign. The Rashba SOC could possibly be made constant on larger domains by external fields perpendicular to graphene. Altogether the valley-Zeeman DW states may form a tunable network of propagating states between domains\cite{SanJose13,Efimkin18,Alsharari16,Huang18,Yoo19,Koshino20}. For untwisted graphene on transition metal dichalcogenide substrates the Moire pattern is on the nanoscale, which might allow the propagating states to remain well-defined, but also might lead to collective effects due to their real-space overlap. Spectroscopic measurements on the nanoscale might be useful to look for the modes, while it would be interesting to expand this study with the effects of local strain due to incommensurability.

The connection between a valley-Zeeman DW and a DW in gating of spinless bilayer graphene implies possibilities to explore valleytronics ideas. Compared to the modes in the spinless bilayer, the valley-Zeeman DW modes are not doubled and they lack any spin-rotation symmetry, hence they should be more resilient to time-reversal breaking. The spinful gated bilayer setup was used in Ref.~\cite{Klinovaja12} to engineer helical modes by magnetic field, and it is an interesting question how the valley-Zeeman DW modes could be manipulated using external fields.

{As we have been focusing on modes inside the bulkgap, we note that a DW which is smooth on the nanoscale should also host gapped excited modes beside the topologically protected gapless modes\cite{Tchoumakov17,VPberg}. Such gapped SOC DW modes would be observable inside the bulkgap if the DW profile was smooth enough, e.g., for $|\lambda|\sim10$meV having width $70$nm.}\footnote{{This estimate is obtained from an analytical formula for the tower of excited states in the example of Kane-Mele SOC with a very smooth DW profile $\eta(x)$, which in a large region of space is well approximated by a linear profile\cite{Tchoumakov17}. We obtain two series of two-fold degenerate states: $E_n(q_y)=\pm\left\{\sqrt{2v_F\frac{\lambda_0}{l}n + v_F^2q_y^2},\sqrt{2v_F\frac{\lambda_0}{l}(n+1) + v_F^2q_y^2}\right\}$, where $n=0,1,2\ldots$, the slope of the DW profile in real space is $\frac{\lambda_0}{l}\equiv \left. \frac{\mathrm{d}\lkm}{\mathrm{d}x}\right|_{x_0}$, with $x_0$ being defined by $\lkm(x_0)=0$. For typical DW profiles the value $\lambda_0=\lkm(+\infty)-\lkm(-\infty)$. In total, the level at $n=0$ is two-fold degenerate (gapless topological modes), and all others are four-fold, which we confirmed in a tight-binding model. The number of DW excited modes inside the bulkgap is estimated as $n_{max}\sim\frac{|\lkm(\pm\infty)|}{t}\frac{l}{a}$.}}
{Hence, this phenomenology might be relevant for twisted incommensurate heterostructures with large Moire periods. Consequently, a valley-Zeeman DW might host an in-gap tower of propagating modes which could be tuned by changing the spatial variation lengthscale and/or the strength of Rashba SOC. The same could be relevant for Kane-Mele SOC DWs, whose topological modes are gapped by Rashba SOC, but the induced amplitude of Rashba SOC in graphene heterostructures seems to be small enough so that all the modes of a Kane-Mele DW could still be within the bulkgap.}

{In the broader perspective on topological modes in graphene due to domain walls in spin-orbit coupling, one may note that the Kane-Mele SOC $\lkm$ provides bulk-index protected DW modes: either protected by time-reversal symmetry and the $Z_2$ bulk-index (on a topological insulator edge, where $|\lkml|<|\msem|<|\lkmr|$), or protected by $S_z$ spin-rotation symmetry and the bulk spin-Chern number (on a sign-changing DW in $\lkm$ with $|\msem|<|\lkml|,|\lkmr|$),} raising the question whether there are DWs in $\lkm$ which break $S_z$ by design but whose modes are not equivalent to topological insulator edge modes.
A natural candidate is a DW across which the spin-axis of the Kane-Mele coupling rotates. From a different viewpoint, the idea that a spiralling magnetic coupling emulates a constant spin-orbit coupling\cite{Braunecker13} has been fruitful in designing topological modes, so it is natural to ask what modes are associated with a spiralling spin-orbit coupling. {We will show elsewhere that there are no isolated domain-wall-bound modes, but instead they are tied to the bulk modes.}

More generally, our work should motivate further theoretical study of topological defects in spin-orbit coupling, since a domain wall is just the simplest one-dimensional example, while zero-dimensional defects in spin-orbit coupling were also confirmed to host interesting bound states\cite{Sau10,Menard19}. {Methodically, the spectral flow theorem proved useful and informative in understanding both single-valley and two-valley systems with one-dimensional defects in two spatial dimensions because the resulting three-dimensional parameter space had point- and line-like degeneracies in our models. It would be interesting to further apply this approach to the numerous quantum condensed matter models in this category.}

\acknowledgments
We would like to thank P. Delplace for useful discussions.

\appendix

\section{Spectral flow theorem reviewed on example of domain walls in Kane-Mele SOC and Dirac mass}
\label{app:SFT}
\begin{figure}
\includegraphics[width=0.45\textwidth]{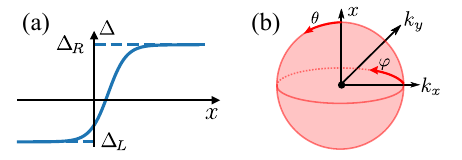}
\caption{\label{fig:app1} Spectral flow theorem for domain walls in the continuum. (a) Generic profile of a continuum DW with arbitrary asymptotic values $\D_L$, $\D_R$ on the two domains. (b) A degeneracy point in classical variable space is enclosed by a sphere $S^2(\theta,\phi)$ on which one calculates the Chern number of the projector to states below the gap.}
\end{figure}
The spectral flow theorem is presented in mathematical detail in Refs.~\cite{Faure19,Delplace20,DelplaceNotes}, while here we demonstrate its practical use in two dimensions with one spatially dependent coupling by solving step-by-step the DW in Kane-Mele coupling $\D(x)\equiv\lkm(x)$ in presence of a constant mass term $\msem$. This example unifies two physical situations as discussed below, and we choose it because each step in the calculation is fully analytical. The theorem involves a few steps:
\begin{enumerate}
\item We replace the quantum Hamiltonian of Eq.~\eqref{eq:Hcont}, $H(x,-i\partial_x,k_y)$, with a matrix function $\tilde{H}(x,k_x,k_y)=k_x\tz\sx+k_y\sy+\msem\sz+\D(x)\tz\sz\spz$ of classical variables so that $k_x$ now commutes with $x$ and thereby with $\eta(x)$ and $\D(x)\equiv\lkm(x)$.
\item We consider $\tilde{H}$ which have some gap in the spectrum that may close at most at some isolated points $(x^i,k_x^i,k_y^i)$ labeled by an integer $i$. Physically, this requires that on both domains far away from the DW the $H$ has a gap between, say the bands $n$ and $n+1$, and we consider the lower $n$ bands filled. The spectrum of our $\tilde{H}$ is $E_{\taus\spins\pm}=\pm\sqrt{k_x^2+k_y^2+(\taus\spins\D(x)+\msem)^2}$, with $\taus=\pm$, $\spins=\pm$ being the eigenvalues of $\tz$, $\spz$, respectively. We are interested in the gap opened at the Dirac point, i.e., half filling, so there are 4 filled bands $E_{\taus\spins-}$.
\item We identify at most two possible degeneracy points closing the gap between filled and empty bands, namely $(x^i,k_x^i,k_y^i)=(x_c^i,0,0)$ where $\taus\spins\D(x_c^i)+\msem=0$ so that
\begin{equation}
  \label{eq:lkmdeg}
  \lkm(x_c^i)=-\taus\spins\msem \in\{+m,-m\},
\end{equation}
and for our smooth monotonous DW profile $\eta(x)$ the equation over all $\taus$, $\spins$ has either:
  \begin{enumerate}
  \item No solutions for $|\D_L|,|\D_R|<|\msem|$ (both domains trivially gapped with the same mass),
  \item One solution if $|\D_L|<|\msem|<|\D_R|$ or $|\D_L|>|\msem|>|\D_R|$ (edge between TI and trivial mass gap),
    \item Both solutions for $|\D_L|,|\D_R|>|\msem|$ with $\text{sgn}(\D_L\D_R)<0$ (DW between two TI domains of opposite sign of $\lkm$).
  \end{enumerate}
\item We enclose a given degeneracy point $(x^i,k_x^i,k_y^i)$ with a closed surface, e.g., sphere $S^i(\theta,\phi)$, and on this surface consider the projector $\tilde{P}^i_-$ onto the filled bands of $\tilde{H}$. For a degeneracy line we need to adapt the enclosing surface. This is detailed in the next  Appendix~\ref{app:restDWs}. The spectral flow theorem states that the number of chiral modes $N_{chiral}^i$ traversing the gap (more precisely, leaving the valence band) is:
  \begin{equation}
    \label{eq:spectralflow}
N^i_{chiral}=C^i_{-},
  \end{equation}
where the chiral modes of positive(negative) velocity along $y$ are counted positively(negatively), and $C^i_{-}$ is the first Chern number of the filled states on the sphere $S^i(\theta,\phi)$:
\begin{equation}
  \label{eq:chern}
  C^i_-=-\frac{1}{2\pi i}\int\limits_0^{2\pi}\!\textrm{d}\phi \!\! \int\limits_0^{\pi}\!\textrm{d}\theta\, \text{Tr} P^i_{-}(\partial_\theta P^i_{-}\partial_\phi P^i_--\partial_\phi P^i_{-}\partial_\theta P^i_{-}).
\end{equation}
The enclosing sphere $S^i(\theta,\phi)$ can be parametrized using $(x,k_x,k_y)=(x^i,k^i_x,k^i_y)+\epsilon (\delta x,\delta k_x,\delta k_y)$ with $(\delta x,\delta k_x,\delta k_y)\equiv(\cos(\theta),\sin(\theta)\cos(\phi),\sin(\theta)\sin(\phi))$, where it is only important to preserve the orientations of the coordinate systems, see Fig.~\ref{fig:app1}b. The coupling term is also approximated on the sphere using the smoothness and monotonicity of the profile $\eta(x)$: $\taus\spins\D(x)+\msem\approx\taus\spins\,D_{RL}\delta x$, $D_{RL}\equiv \text{sgn}(\lkmr-\lkml)$ for the two possible degeneracy points, where we rescaled the coupling by a positive constant which does not change the topology of $\tilde{P}$  (similarly we set the radius $\epsilon=1$). The projectors to the 4 filled bands are
\begin{align}
  \label{eq:proj2x2}
  &P^i_{\taus\spins,-}=\frac{1}{2}(\id-\hat{d}^i_{\taus,\spins}\cdot\sigmas)\\
  &\hat{d}^i_{\taus,\spins}=(\taus\sin(\theta)\cos(\phi),\sin(\theta)\sin(\phi),\taus\spins D_{RL}\cos(\theta)).
\end{align}
The standard Chern number of spin-$1/2$ in magnetic field implies that $C_{\taus\spins,-}=s D_{RL}$, since $C$ is preserved under inversion of $\hat{d}$ but flips sign under mirror operations. A band contributes one mode ($|C_{\taus\spins,-}|=1$) for each degeneracy point its $\taus$, $\spins$ give according to Eq.~\eqref{eq:lkmdeg}. In particular, if $\lkmr>\lkml$ then a Kane-Mele DW profile $\lkm(x)$ which crosses the value $m$ hosts two chiral modes (Kramers pair) with quantum numbers $\taus=-\spins$, while if it crosses the value $-m$ there are two (more) modes with $\taus=\spins$, with chiralities always $N_{chiral}=\spins$.
\end{enumerate}

As expected we recover the Kramers pair of topological insulator edge modes, and we find the four modes of a Kane-Mele DW across which $\lkm$ changes sign (while $|\msem|<|\lkmr|,|\lkml|$, possibly $\msem=0$), as expected from the spin Chern number difference of $2$.

Using the above approach we also find the well-known valley-polarized modes for a DW with a sign change of mass $\msem$ (given that $|\lkm|<|\msem_R|,|\msem_L|$).\cite{Semenoff08,Faure19}

\section{Absence of topological modes for other DWs involving valley-Zeeman and Rashba SOC}
\label{app:restDWs}
We start by considering a DW in either $\lvz$ or in $\lr$ in presence of a constant $\msem$, and find no topological DW modes. Concretely, in the first case the spectrum of $\tilde{H}$ is $E_{\taus\spins}=\taus\spins\lvz\pm\sqrt{k^2+\msem^2}$, which is either gapless or has no degeneracy point, while in the second case $E_{\alpha\beta}=\alpha\sqrt{k^2+\msem^2+2\lr^2+2\beta|\lr|\sqrt{k^2+\lr^2}}$, with $\alpha,\beta=\pm$, has no degeneracy points for any value of $\lr(x)$.

More interestingly, we now consider a DW in $\D(x)\equiv\lr(x)$ in presence of constant $\lvz\neq0$. Note that the bulkgap of graphene with both constant $\lr,\lvz$ is equal to $\frac{2|\lr\lvz|}{\sqrt{\lr^2+\lvz^2}}$, which is an expression symmetric to exchange of $\lvz$ and $\lr$. Nevertheless the DWs are completely different: the DW in $\lvz$ with constant $\lr$ hosts robust modes discussed in Section~\ref{sec:secvzDW}, while in the following we show that a DW in $\lr$ with constant $\lvz$ hosts no modes at all. A DW in $\D(x)\equiv\lr(x)$ in presence of constant $\lvz\neq0$ implies that a degeneracy at half-filling occurs on a ring
\begin{equation}
  \label{eq:1b}
  (x^\phi,k_x^\phi,k_y^\phi)=(x_c,|\lvz|\cos(\phi),|\lvz|\sin(\phi)),
\end{equation}
where $\lr(x_c)=0$, Therefore propagating modes are {\emph a priori} possible only if the Rashba SOC changes sign across the DW. The spectral flow theorem dictates that we enclose the entire line of degeneracy points with a surface, in this case the ring is simply enclosed by a torus of small radius $\epsilon$:
\begin{equation}
  \label{eq:2}
  \begin{pmatrix}
    \lr\\
    k_x\\
    k_y
  \end{pmatrix}=
  \begin{pmatrix}
    \epsilon\cos(\theta)\\
    (|\lvz|+\epsilon\sin(\theta))\cos(\phi)\\
    (|\lvz|+\epsilon\sin(\theta))\sin(\phi)
  \end{pmatrix}
\end{equation}
The Chern number of the projector onto the filled bands over the torus surface (one just changes the range of $\theta\in[0,2\pi]$ in Eq.~\eqref{eq:chern}) gives zero, and therefore there are no topological gapless modes on the Rashba DW. We confirmed the absence of topological DW modes traversing the bulkgap on the lattice too, for both smooth and sharp DW profiles.

\section{Exact solution of domain walls in Kane-Mele and Dirac mass couplings}
\label{app:exact}
{Here we demonstrate that an exact solution for zero-modes of for arbitrary Dirac mass and Kane-Mele SOC domain walls in the continuum matches the predictions of the SFT. It is obvious that this calculation is tedious, opaque and requires ansatzes.}

Given the rotation invariance of the low-energy Hamiltonian we can consider a domain wall in the $x$ direction without loss of generality. We start with Eq.~(\ref{eq:Hcont}) with the usual substitution $k_x \rightarrow -i\partial_x$:
\begin{equation}
	\mathcal{H} = -iv_F\partial_x\tau_z\sigma_x + v_Fk_y\sigma_y + \lambda (x)\tau_zs_z\sigma_z + m(x)\sigma_z~.
\end{equation}

We rewrite the eigenproblem $\mathcal{H}\Psi = E\Psi$ as an explicit first order differential equation :
\begin{equation}
\partial_x\Psi = \underbrace{\left(k_y\tau_z\sigma_z - \frac{\lambda (x)}{v_F}s_z\sigma_y - \frac{m(x)}{v_F}\tau_z\sigma_y +i\frac{E}{v_F}\tau_z\sigma_x\right)}_{\displaystyle\Lambda (x)}\Psi
\end{equation}

Since $\Lambda$ is block diagonal in the valley and spin space, we can compute the exponential in the different eigenspace independently. The general solution requires to compute the space ordered integral $\exp(\int_0^x \Lambda(t)\mathrm{d}t)$. Since we are looking for zero modes,we restrict ourselves to $k_y=0$ and $E=0$ which makes the exponential easy to obtain. Doing so we will lose information on the chirality, but we will come back to it at the end. We introduce the following intermediate notations:

\begin{equation*}
	a(x) = \frac{1}{v_F}\int_0^x s^z\lambda(t) + \xi m(t) \,\mathrm{d}t
\end{equation*}
so that $\int_0^x \Lambda (t)\,\mathrm{d}t = -a(x)\sy$ where $\xi$ and $s^z$ are respectively the eigenvalues of $\tau_z$ and $s_z$. As a consequence
\begin{equation}
	\exp(-a(x)\sy) = \mathrm{ch}(a(x))\id - \mathrm{sh}(a(x))\sy~.
\end{equation}
Now we look for solutions of the differential equation which are normalizable. Let us assume that $\lambda$ and $m$ have finite limits in $\pm \infty$. Consequently

\begin{align*}
	a(x) &\underset{+\infty}{\sim} \frac{1}{v_F}x(s^z\lambda_{+\infty}+\xi m_{+\infty}) = xa_+ \\
	a(x) &\underset{-\infty}{\sim} \frac{1}{v_F}x(s^z\lambda_{-\infty}+\xi m_{-\infty}) = xa_-~.
\end{align*}

This implies that
\begin{equation}
	\exp(a\sy) \underset{\pm\infty}{\sim} \frac{\exp(|a_{\pm} x|)}{2}\id - \mathrm{sgn}(xa_{\pm})\frac{\exp(|a_{\pm}x|)}{2}\hat{n}.\vec{\sigma}~.
\end{equation}

To be normalizable, the divergent components of $\exp(a(x)\sy)$ must be simultaneously zero at $\pm\infty$, which means that
\begin{equation}
	\ker(\id-\mathrm{sgn}(a_+)\sy) \cap \ker(\id+\mathrm{sgn}(a_-)\sy) \neq \{0\}~.
\end{equation}
This is true if and only if $a_+a_-<0$.

If $m+\lambda$ changes sign, then we have a Kramers' pair with $\xi=s^z$. If $m-\lambda$ changes sign then we have an other one with $\xi=-s^z$.

Chirality can be most simply recovered in the limit of a wide DW, and we do not expect that chirality is flipped under smooth local deformations of the DW profile, including collapsing the DW into a discontinuous step-like potential. This limiting case can be fully solved with the previous method, albeit with more difficulties than for a smooth DW. Hence we will linearize the DW around a position where $\Delta=m+\xi s^z \lambda$ changes sign. We already saw a change of sign in $\Delta$ is a sufficient and necessary condition for the zero modes.
\begin{equation}
	\Delta = \left.\frac{\mathrm{d}\Delta(x)}{\mathrm{d}x}\right|_{x_0}(x-x_0)=\frac{\Delta_0}{l}(x-x_0)\,,
\end{equation}
where $x_0$ is the position where $\Delta$ goes to 0. Let's first focus on $k_y=0$. By squaring the Hamiltonian we obtain an harmonic oscillator and an homogeneous term :
\begin{equation}
	\mathcal{H}^2 = -v_F^2\partial_x^2+\frac{\Delta_0^2}{l^2}(x-x_0)^2-v_F\frac{\xi\Delta_0}{l}\sy\,.
	\label{eq:appA_HO}
\end{equation}
The eigenenergies of the harmonic oscillator part are $2v_F|\Delta_0/l|(n+1/2)$ with $n$ a positive or null integer. The last term of Eq.~\ref{eq:appA_HO} thus precisely shifts these energies to $2v_F|\Delta_0/l|n$ and $2v_F|\Delta_0/l|(n+1)$. In particular the zero energy subspace is not degenerate as opposed to all other states which are two-fold degenerate and which contain both eigenvectors of $\sy$. So, the zero energy eigenstate is also a eigenstate of $\sy$ with eigenvalue $\xi\mathrm{sgn}(\Delta_0)$. And of course, if $m+\xi s^z\lambda$ changes sign, then $m+(-\xi)(-s^z)\lambda$ also changes sign in the same way.

Now, since $\mathcal{H}^2=\mathcal{H}^2(k_y=0)+v_F^2k_y^2$, this means that the zero eigenvector of $\mathcal{H}(k_y=0)$ has eigenvalue $\sy v_Fk_y$ where we identify $\sy$ with its eigenvalue. So, we recapitulate the chiralities of the different modes in the different cases in the following table :

\begin{figure}[h]
\begin{tabular}{|c|c|c|c|}
	\hline
	\multicolumn{2}{|c|}{$m+\lambda$} & \multicolumn{2}{c|}{$m-\lambda$} \\
	\hline
	$\Delta_0>0$ & $\Delta_0<0$ & $\Delta_0>0$ & $\Delta_0<0$ \\
	\hline
	$+\,\uparrow$ RM & $-\,\downarrow$ RM & $+\,\downarrow$ RM & $-\,\uparrow$ RM \\
	$-\,\downarrow$ LM & $+\,\uparrow$ LM & $-\,\uparrow$ LM & $+\,\downarrow$ LM \\
	\hline
\end{tabular}
\end{figure}
The first row denotes which term is changing sign while $\pm$ is the value of $\xi$, $\uparrow\downarrow$ the value of $s^z$ and RM (LM) means right (left) mover.For a DW between a trivial and QSH phase, then either $m+\lambda$ or $m-\lambda$ changes sign, but they cannot both change sign. We recover the standard edge states with direction of movement, spin and valley being correlated. If both terms change sign, then the DW separates either two trivial or two QSH phases. If it separate two trivial phases, then both $m+\lambda$ and $m-\lambda$ have the same sign. Thus, we have both spins at both valley, but valley and direction of movement are still correlated. If it separates two QSH phases, then $m+\lambda$ and $m-\lambda$ have opposite signs, so direction of movement is now only correlated with spin.

These match exactly the results of applying the spectral flow theorem.

\section{Intervalley-scattering gap for armchair domain wall in Dirac mass}
\label{app:mass}
{Here we derive an analytical perturbative result for the small gap due to intervalley mixing of valley-polarized modes of a DW in Dirac mass.}

Continuum theory (either SFT or direct solution) predicts four in-gap modes and hence four zero-energy states, with both spins present in both valleys. Here we consider an armchair domain wall in the mass term and perturbatively compute the valley-mixing effect near the Dirac momenta. Our initial eigenspace is the four-fold degenerate zero energy eigenspace for spinless fermions (so eight-fold for electrons) of standard graphene and our perturbation is the mass domain wall:
\begin{widetext}
\begin{equation}
  H_{\mathrm{trivial}} = \displaystyle\sum_{\beta=0}^{\frac{N_y}{2}-1}\sum_{\alpha=0}^{N_x-1} m_\alpha(c^\dagger_{\alpha\beta,A1}c_{\alpha\beta,A1} + c^\dagger_{\alpha\beta,A2}c_{\alpha\beta,A2} - c^\dagger_{\alpha\beta,B1}c_{\alpha\beta,B1}-c^\dagger_{\alpha\beta,B2}c_{\alpha\beta,B2}),~\text{where}~
  m_\alpha =
	\begin{cases}
		+m~\text{if~}\alpha<\frac{Nx}{2} \\
		-m~\text{otherwise}
	\end{cases}
      \end{equation}
    \end{widetext}
Here we introduce an enlarged unit-cell comprising four atoms which belong to two adjacent elementary unit-cells. We use the numbers $\{1,2\}$ to label this internal degree of freedom. This folds the hexagonal Brillouin zone into a rectangle where $\pm\mathbf{K}$ are mapped to $\mp 2\pi/(3\sqrt{3}a)$.

The first step is to compute the matrix elements of the mass term in the Fourier basis.
\begin{widetext}
\begin{equation}
	H_\mathrm{trivial} = \displaystyle\sum_{k_y=0}^{\frac{N_y}{2}-1}\sum_{\substack{k_1=0 \\ k_2=0 \\ k_1-k_2 \equiv 1[2]}}^{N_x-1} m\frac{1}{N_x}\frac{4}{1-e^{2i\pi \frac{(k_1-k_2)}{N_x}}}(c^\dagger_{k_1k_y,A1}c_{k_2k_y,A1} + c^\dagger_{k_1k_y,A2}c_{k_2k_y,A2}
-c^\dagger_{k_1k_y,B1}c_{k_2k_y,B1}-c^\dagger_{k_1k_y,B2}c_{k_2k_y,B2})
\end{equation}
\end{widetext}
Because we considered a sharp domain wall and symmetric domains we only couple states with an odd momentum difference. As a consequence there is no first order contribution of the mass on the Dirac cones, so we need to go to second order. To do so we need the eigenvectors with energies close to 0 which we obtain by using first order perturbation theory on the zero energy eigenspace of standard graphene but the perturbation is now the small momentum in the $\vec{x}$ direction.

The Bloch Hamiltonian at the Dirac point $\mathbf{K} = \frac{N_x}{3} \vec{e_x}$ is, in units $t=-1$:
\begin{equation}
	H_\mathbf{K}=\begin{pmatrix}
		0 & 0 & 1 & -j^2 \\
		0 & 0 & -j & 1 \\
		1 & -j^2 & 0 & 0 \\
		-j & 1 & 0 & 0
	\end{pmatrix},
\end{equation}
where $j=e^{\frac{2i\pi}{3}}$ is the usual cubic root of 1, and the Bloch Hamiltonian $H_{-\mathbf{K}}$ at the other valley $-\mathbf{K}$ is obtained simply by exchanging $j$ and $j^2$. Now we can compute the effect at first order, namely at first order in $\frac{p}{N_x}$,

\begin{equation}
	H_{\mathbf{K}+p\vec{e_x}} = H_\mathbf{K}+\underbrace{\frac{2i\pi p}{N_x}\begin{pmatrix}
		0 & 0 & 0 & j \\
		0 & 0 & -j^2 & 0 \\
		0 & j & 0 & 0 \\
		-j^2 & 0 & 0 & 0
	\end{pmatrix}}_{V}~.
\end{equation}

\begin{figure}
\includegraphics{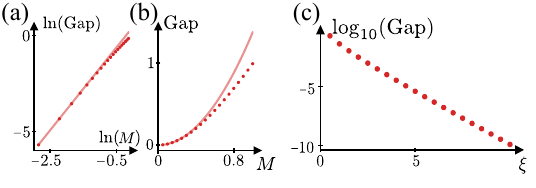}
\caption{\label{fig:app2} Intervalley scattering in a tight-binding lattice model for a sign-changing armchair DW in Dirac mass, with parameters $N_x=204$, $N_y=500$, $\msem(\pm\infty)=\pm0.1t$. (a), (b) The size of the gap at $k_y=0$ opened in the valley-polarized DW modes due to intervalley scattering of an atomically sharp DW profile. Measured in tight-binding (dots) and derived perturbatively (line) as function of $M\equiv|\msem(x\rightarrow\pm\infty)/t|$, in log-log scale (a) and lin-lin scale (b). (c) Exponential decay of the gap in DW modes (in units of $t$) as a function of the width of the domain wall (in units of lattice constant).}
\end{figure}

Details of the calculation are not presented for the sake of brevity, and finally we obtain that the effect of the mass term at second order on the low energy states of graphene in the $\{\frac{j^2\ket{A_1, \mathbf{K}}+\ket{A_2, \mathbf{K}}}{\sqrt{2}},$ $\frac{j^2\ket{B_1, \mathbf{K}}+\ket{B_2, \mathbf{K}}}{\sqrt{2}}, \frac{j\ket{A_1, -\mathbf{K}}+\ket{A_2, -\mathbf{K}}}{\sqrt{2}}, \frac{j\ket{B_1, -\mathbf{K}}+\ket{A_2, -\mathbf{K}}}{\sqrt{2}}\}$ basis is :

\begin{equation}
	H^{(2)} = \begin{pmatrix}
	0 & 0 & 0 & \kappa^* \\
	0 & 0 & \kappa^* & 0 \\
	0 & \kappa & 0 & 0 \\
	\kappa & 0 & 0 & 0
\end{pmatrix} = (\Re(\kappa)\tau_x+\Im(\kappa)\tau_y)\sigma_x,
\end{equation}
where
\begin{align}\notag
\kappa &= \frac{16m^2}{N_x^2}\displaystyle\sum_{p\in\mathbb{Z}}\frac{1}{(1-e^{2i\pi\frac{2p+1}{N_x}})(j-e^{-2i\pi\frac{2p+1}{N_x}})}\frac{1}{v_x(2p+1)},\\
v_x &= \frac{\sqrt{3}\pi}{N_x}.
\end{align}
This Hamiltonian opens a $2|\kappa|$ gap. To compute $\kappa$ we use that $N_x\rightarrow+\infty$, hence
\begin{equation}
\kappa \approx \frac{16m^2}{N_x^2}\displaystyle\sum_{p\in\mathbb{Z}}\frac{1}{2i\pi\frac{2p+1}{N_x}(1-j-2i\pi\frac{2p+1}{N_x})}\frac{1}{v_x(2p+1)}.
\end{equation}
Since $v_x$ is in $\frac{1}{N_x}$ it is justified to only keep the first order contribution in $\frac{p}{N_x}$, which will lead to a constant term thanks to the prefactor $\frac{1}{N_x^2}$, while higher order contributions will disappear in the thermodynamic limit. We get
\begin{align}
	\kappa &\approx \frac{16m^2}{(1-j)N_x^2}\displaystyle\sum_{p\in\mathbb{Z}}\frac{N_x^2}{2\sqrt{3}i\pi^2(2p+1)^2}\\ \nonumber
	&= \frac{8m^2}{i(1-j)\sqrt{3}\pi^2} \sum_{p\in\mathbb{Z}} \frac{1}{(2p+1)^2} = \frac{8m^2}{i(1-j)\sqrt{3}\pi^2}2\frac{\pi^2}{8}\\
        &= -\frac{2m^2}{\sqrt{3}(1-j)}i,
\end{align}
and finally
\begin{equation}
	|\kappa|=\frac{2m^2}{3},
\end{equation}
which is the expression quoted in the main text. The predicted gap of $2|\kappa|$ is compared to the gap in modes we obtain by numerical exact diagonalization of the tight-binding Hamiltonian (Fig.~\ref{fig:app2}a,b).

Finally if we go from a sharp domain wall to a smooth domain wall, the gap gets drastically reduced, as can be seen in Fig.~\ref{fig:app2}c. Our smooth domain walls are obtained by convolution of a sharp DW (step function) with a gaussian of desired width $\xi$. The drop in the value of the gap is coherent with our analysis as the value of the gap is tuned by the Fourier coefficients linking states close to $K$ to states close to $-K$. Indeed, as the smoothness of the domain wall increases, which means as its typical width increases, the width of its Fourier transform decreases. This gives support to our assessment that the gap is a second order effect caused by the sharpness of the domain wall.

\bibliography{SO_DWrefs}
\end{document}